\input harvmac



\lref\bhk{ M.~Berg, M.~Haack and B.~K\"ors, ``On the Moduli
Dependence of Nonperturbative Superpotentials in Brane
Inflation,'' hep-th/0409282.
}

\lref\bhko{ M.~Berg, M.~Haack and B.~K\"ors, ``Loop Corrections to
Volume Moduli and Inflation in String Theory,'' Phys.\ Rev.\ D
{\bf 71}, 026005 (2005) hep-th/0404087.
}

\lref\pisin{
P.~Chen, K.~Dasgupta, K.~Narayan, M.~Shmakova and M.~Zagermann,
``Brane inflation, solitons and cosmological solutions: I,''
hep-th/0501185.
}

\lref\kklt{ S.~Kachru, R.~Kallosh, A.~Linde and S.~P.~Trivedi,
``de Sitter vacua in string theory,'' Phys.\ Rev.\ D {\bf 68},
046005 (2003), hep-th/0301240.
}

\lref\inflation{ A.D. Linde, {\it{Particle Physics and Inflationary
Cosmology}}, Harwood Academic, 1990\semi
E.W. Kolb and M.S. Turner, {\it{The Early Universe}}, Addison-Wesley,
1990\semi
A.R. Liddle and D.H. Lyth, {\it{Cosmological Inflation and Large-Scale
Structure}}, Cambridge University Press, 2000.}

\lref\DeWolfe{ O.~DeWolfe and S.~B.~Giddings, ``Scales and
hierarchies in warped compactifications and brane worlds,'' Phys.\
Rev.\ D {\bf 67}, 066008 (2003), hep-th/0208123.
}

\lref\intriligator{
K.~A.~Intriligator and N.~Seiberg,
``Lectures on supersymmetric gauge theories and electric-magnetic  duality,''
Nucl.\ Phys.\ Proc.\ Suppl.\  {\bf 45} BC, 1 (1996), hep-th/9509066.
}

\lref\BuchelQJ{
A.~Buchel and R.~Roiban,
``Inflation in warped geometries,''
Phys.\ Lett.\ B {\bf 590}, 284 (2004),
hep-th/0311154.
}

\lref\dvali{ G. Dvali, personal communication.}

\lref\gsw{ M.B. Green, J.H. Schwarz, and E. Witten, {\it{Superstring Theory, Volume 2}},
Cambridge, 1986. }

\lref\lukas{ B.~de Carlos, A.~Lukas and S.~Morris,
``Non-perturbative vacua for M-theory on G(2) manifolds,''
hep-th/0409255.
}

\lref\shifman{M.~A.~Shifman and A.~I.~Vainshtein, ``On Gluino
Condensation In Supersymmetric Gauge Theories. SU(N) And O(N)
Groups,'' Nucl.\ Phys.\ B {\bf 296}, 445 (1988)\semi M.~A.~Shifman
and A.~I.~Vainshtein, ``On holomorphic dependence and infrared
effects in supersymmetric gauge theories,'' Nucl.\ Phys.\ B {\bf
359}, 571 (1991).
}

\lref\BeckerNN{ K.~Becker, M.~Becker, M.~Haack and J.~Louis,
``Supersymmetry breaking and alpha'-corrections to flux induced
potentials,'' JHEP {\bf 0206}, 060 (2002), hep-th/0204254.
}

\lref\louis{ V.~Kaplunovsky and J.~Louis, ``On gauge couplings in
string theory,'' Nucl.\ Phys.\ B {\bf 444}, 191 (1995),
hep-th/9502077\semi
J.~Louis and K.~Foerger, ``Holomorphic couplings in string
theory,'' Nucl.\ Phys.\ Proc.\ Suppl.\  {\bf 55} B, 33 (1997),
hep-th/9611184.
}

\lref\hetinf{ E.~I.~Buchbinder, ``Five-brane dynamics and
inflation in heterotic M-theory,'' hep-th/0411062\semi
K.~Becker, M.~Becker and A.~Krause, ``M-Theory Inflation from
Multi M5-Brane Dynamics,'' hep-th/0501130.
}

\lref\GVW{ S.~Gukov, C.~Vafa and E.~Witten, ``CFT's from
Calabi-Yau four-folds,'' Nucl.\ Phys.\ B {\bf 584}, 69 (2000),
hep-th/9906070.
}

\lref\Witten{ E.~Witten, ``Non-Perturbative Superpotentials In
String Theory,'' Nucl.\ Phys.\ B {\bf 474}, 343 (1996),
hep-th/9604030.
}

\lref\GKP{ S.~B.~Giddings, S.~Kachru and J.~Polchinski,
``Hierarchies from fluxes in string compactifications,'' Phys.\
Rev.\ D {\bf 66}, 106006 (2002), hep-th/0105097.
}

\lref\sorokin{
R.~Kallosh and D.~Sorokin,
``Dirac action on M5 and M2 branes with bulk fluxes,''
hep-th/0501081.
}

\lref\eucl{ L.~G\"orlich, S.~Kachru, P.~K.~Tripathy and
S.~P.~Trivedi, ``Gaugino condensation and nonperturbative
superpotentials in flux compactifications,'' hep-th/0407130.
}

\lref\angel{ C.~Angelantonj, R.~D'Auria, S.~Ferrara and
M.~Trigiante, ``K3 x T**2/Z(2) orientifolds with fluxes, open
string moduli and critical points,'' Phys.\ Lett.\ B {\bf 583},
331 (2004), hep-th/0312019.
}

\lref\watari{ F.~Koyama, Y.~Tachikawa and T.~Watari,
``Supergravity analysis of hybrid inflation model from D3-D7
system,'' Phys.\ Rev.\ D {\bf 69}, 106001 (2004), hep-th/0311191.
}

\lref\kl{ R.~Kallosh and A.~Linde, ``Landscape, the scale of SUSY
breaking, and inflation,'' hep-th/0411011.
}

\lref\DasguptaDW{
K.~Dasgupta, J.~P.~Hsu, R.~Kallosh, A.~Linde and M.~Zagermann,
``D3/D7 brane inflation and semilocal strings,''
JHEP {\bf 0408}, 030 (2004), hep-th/0405247.
}

\lref\hetero{
E.~I.~Buchbinder and B.~A.~Ovrut,
``Vacuum stability in heterotic M-theory,''
Phys.\ Rev.\ D {\bf 69}, 086010 (2004), hep-th/0310112\semi
S.~Gukov, S.~Kachru, X.~Liu and L.~McAllister, ``Heterotic moduli
stabilization with fractional Chern-Simons invariants,'' Phys.\
Rev.\ D {\bf 69}, 086008 (2004), hep-th/0310159\semi
M.~Becker, G.~Curio and A.~Krause, ``De Sitter vacua from
heterotic M-theory,'' Nucl.\ Phys.\ B {\bf 693}, 223 (2004),
hep-th/0403027.
}

\lref\angles{
J.~Garcia-Bellido, R.~Rabadan and F.~Zamora, ``Inflationary
scenarios from branes at angles,'' JHEP {\bf 0201}, 036 (2002),
hep-th/0112147.
}

\lref\pseudo{
S.~Buchan, B.~Shlaer, H.~Stoica and S.~H.~H.~Tye, ``Inter-brane
interactions in compact spaces and brane inflation,'' JCAP {\bf
0402}, 013 (2004), hep-th/0311207.
}

\lref\forthcoming{M. Berg, M. Haack, and B. K\"ors, work in
progress.}

\lref\Halyo{ E.~Halyo, ``Hybrid inflation from supergravity
D-terms,'' Phys.\ Lett.\ B {\bf 387}, 43 (1996), hep-ph/9606423.
}

\lref\Kallosh{K. Dasgupta, C. Herdeiro, S. Hirano and R. Kallosh,
``D3/D7 Inflationary Model and M-theory,'' Phys. Rev. D {\bf 65}, 126002
(2002), hep-th/0203019.}

\lref\BinetruyXJ{ P.~Binetruy and G.~R.~Dvali, ``D-term
inflation,'' Phys.\ Lett.\ B {\bf 388}, 241 (1996),
hep-ph/9606342. }

\lref\alexone{ A.~Saltman and E.~Silverstein, ``The scaling of the
no-scale potential and de Sitter model building,'' JHEP {\bf
0411}, 066 (2004), hep-th/0402135.
}

\lref\DDF{ F.~Denef, M.~R.~Douglas and B.~Florea, ``Building a
better racetrack,'' JHEP {\bf 0406}, 034 (2004), hep-th/0404257.
}

\lref\BKQ{ C.~P.~Burgess, R.~Kallosh and F.~Quevedo, ``de Sitter
string vacua from supersymmetric D-terms,'' JHEP {\bf 0310}, 056
(2003), hep-th/0309187.
}

\lref\Copeland{ E.~J.~Copeland, A.~R.~Liddle, D.~H.~Lyth,
E.~D.~Stewart and D.~Wands, ``False vacuum inflation with Einstein
gravity,'' Phys.\ Rev.\ D {\bf 49}, 6410 (1994), astro-ph/9401011.
}

\lref\dt{ M.~Dine, L.~Randall and S.~Thomas, ``Supersymmetry
breaking in the early universe,'' Phys.\ Rev.\ Lett.\  {\bf 75},
398 (1995), hep-ph/9503303.
}

\lref\scott{ S. Thomas, personal communication.}

\lref\others{ S. Alexander, ``Inflation from D - Anti-D-Brane
Annihilation,'' Phys. Rev. D {\bf 65}, 023507 (2002),
hep-th/0105032\semi G. Dvali, Q. Shafi and S. Solganik, ``D-brane
Inflation,'' hep-th/0105203\semi C.P. Burgess, M. Majumdar, D.
Nolte, F. Quevedo, G. Rajesh and R.J. Zhang, ``The Inflationary
Brane-Antibrane Universe,'' JHEP {\bf 07}, 047 (2001),
hep-th/0105204.}

\lref\eight{ J.~J.~Blanco-Pillado {\it et al.}, ``Racetrack
inflation,'' JHEP {\bf 0411}, 063 (2004), hep-th/0406230.
}

\lref\shifman{ M.~A.~Shifman and A.~I.~Vainshtein, ``Solution Of
The Anomaly Puzzle In Susy Gauge Theories And The Wilson Operator
Expansion,'' Nucl.\ Phys.\ B {\bf 277}, 456 (1986).
}

\lref\NillesCY{ H.~P.~Nilles, ``The Role Of Classical Symmetries
In The Low-Energy Limit Of Superstring Theories,'' Phys.\ Lett.\ B
{\bf 180}, 240 (1986).
}

\lref\otherwarped{
C.~P.~Burgess, J.~M.~Cline, H.~Stoica and F.~Quevedo, ``Inflation
in realistic D-brane models,'' JHEP {\bf 0409}, 033 (2004),
hep-th/0403119\semi
N.~Iizuka and S.~P.~Trivedi, ``An inflationary model in string
theory,'' Phys.\ Rev.\ D {\bf 70}, 043519 (2004),
hep-th/0403203\semi
X.~G.~Chen, ``Multi-throat brane inflation,'' hep-th/0408084\semi
X.~G.~Chen, ``Inflation from warped space,'' hep-th/0501184\semi
H.~Firouzjahi and S.~H.~Tye,
``Brane inflation and cosmic string tension in superstring theory,''
hep-th/0501099.
}

\lref\PiSS{ E.~Silverstein, ``TASI / PiTP / ISS lectures on moduli
and microphysics,'' hep-th/0405068.
}

\lref\sandip{ L. G\"orlich, P. Tripathy and S. Trivedi, work in
progress.}

\lref\tt{
P.~K.~Tripathy and S.~P.~Trivedi,
``Compactification with flux on K3 and tori,''
JHEP {\bf 0303}, 028 (2003), hep-th/0301139.
}

\lref\kst{
S.~Kachru, M.~B.~Schulz and S.~Trivedi,
``Moduli stabilization from fluxes in a simple IIB orientifold,''
JHEP {\bf 0310}, 007 (2003), hep-th/0201028.
}

\lref\BinetruyHH{
P.~Binetruy, G.~Dvali, R.~Kallosh and A.~Van Proeyen,
``Fayet-Iliopoulos terms in supergravity and cosmology,''
Class.\ Quant.\ Grav.\  {\bf 21}, 3137 (2004), hep-th/0402046.
}

\lref\sarah{
S.~E.~Shandera,
``Slow Roll in Brane Inflation,''
hep-th/0412077.
}

\lref\supercosmo{ R.~Kallosh and S.~Prokushkin,
``SuperCosmology,'' hep-th/0403060.
}

\lref\alextwo{ A.~Saltman and E.~Silverstein, ``A new handle on de
Sitter compactifications,'' hep-th/0411271.
}

\lref\dvalitye{ G.~R.~Dvali and S.~H.~H.~Tye, ``Brane inflation,''
Phys.\ Lett.\ B {\bf 450}, 72 (1999), hep-ph/9812483.
}

\lref\kklmmt{ S.~Kachru, R.~Kallosh, A.~Linde, J.~Maldacena,
L.~McAllister and S.~P.~Trivedi, ``Towards inflation in string
theory,'' JCAP {\bf 0310}, 013 (2003), hep-th/0308055.
}

\lref\shift{J.~P.~Hsu and R.~Kallosh, ``Volume stabilization and
the origin of the inflaton shift symmetry in string theory,'' JHEP
{\bf 0404}, 042 (2004), hep-th/0402047\semi
H.~Firouzjahi and S.~H.~H.~Tye, ``Closer towards inflation in
string theory,'' Phys.\ Lett.\ B {\bf 584}, 147 (2004),
hep-th/0312020\semi
J.~P.~Hsu, R.~Kallosh and S.~Prokushkin, ``On brane inflation with
volume stabilization,'' JCAP {\bf 0312}, 009 (2003),
hep-th/0311077.
}

\lref\CamaraJJ{
P.~G.~Camara, L.~E.~Ibanez and A.~M.~Uranga,
``Flux-induced SUSY-breaking soft terms on D7-D3 brane systems,''
hep-th/0408036.
}

\lref\LustFI{
D.~Lust, S.~Reffert and S.~Stieberger,
``Flux-induced soft supersymmetry breaking in chiral type IIb orientifolds
with D3/D7-branes,'' hep-th/0406092.
}



\font\cmss=cmss10 \font\cmsss=cmss10 at 7pt

\def\IC{\relax\hbox{$\inbar\kern-.3em{\rm C}$}}
\def\IR{\relax{\rm I\kern-.18em R}}
\def\Z{\relax\ifmmode\mathchoice
{\hbox{\cmss Z\kern-.4em Z}}{\hbox{\cmss Z\kern-.4em Z}}
{\lower.9pt\hbox{\cmsss Z\kern-.4em Z}} {\lower1.2pt\hbox{\cmsss
Z\kern-.4em Z}}\else{\cmss Z\kern-.4em
Z}\fi}

\hskip 1cm

\vskip -1 cm

\Title{\vbox{\baselineskip12pt \hbox{hep-th/0502001}
\hbox{SLAC-PUB-11008} \hbox{SU-ITP-05/05} }} {\vbox{\centerline{An
Inflaton Mass Problem in String Inflation} \centerline{from}
\centerline{Threshold Corrections to Volume Stabilization}}}

\centerline{Liam McAllister}

\medskip
\centerline{\it Department of Physics and SLAC}
\centerline{\it Stanford University, Stanford, CA 94305 USA}

\vskip .2in
\medskip\

 Inflationary models whose vacuum energy arises from a D-term are believed
not to suffer from the supergravity eta problem of F-term
inflation. That is, D-term models have the desirable property that
the inflaton mass can naturally remain much smaller than the
Hubble scale.  We observe that this advantage is lost in models
based on string compactifications whose volume is stabilized by a
nonperturbative superpotential: the F-term energy associated with
volume stabilization causes the eta problem to reappear. Moreover,
any shift symmetries introduced to protect the inflaton mass will
typically be lifted by threshold corrections to the
volume-stabilizing superpotential. Using threshold corrections
computed by Berg, Haack, and K\"ors, we illustrate this point in
the example of the D3-D7 inflationary model, and conclude that
inflation is possible, but only for fine-tuned values of the
stabilized moduli. More generally, we conclude that inflationary
models in stable string compactifications, even D-term models with
shift symmetries, will require a certain amount of fine-tuning to
avoid this new contribution to the eta problem.

\bigskip\
\bigskip\
\bigskip\

\newsec{Introduction}

  In any model of slow-roll inflation \inflation, one needs the inflaton
potential $V(\phi)$ to be rather flat, as measured by the
slow-roll parameters: \eqn\epsis{ \epsilon \equiv {M_{{\rm{p}}}^2\over{2}}
\Bigl({V'\over{V}}\Bigr)^{2}} \eqn\etais{ \eta \equiv M_{{\rm{p}}}^2
\Bigl({V''\over{V}}\Bigr)} where $M_{\rm{p}}$ is the four-dimensional
reduced Planck mass and primes denote derivatives with respect to
the inflaton $\phi$.  It is convenient to rewrite \etais\ as
\eqn\etaalso{ \eta = {V''\over{3H^2}} } so that $\eta$ measures
the inflaton mass in units of the Hubble scale H. Observations
require that $\eta \le 10^{-2}$.  A key issue in inflationary
model-building is the solution of this constraint.

  Inflationary models in supergravity can be divided into F-term models
and D-term models according to the source of the
supersymmetry-breaking energy which drives inflation.  F-term
models suffer from what is known as the eta problem, or the
inflaton mass problem \Copeland.  The F-term energy \eqn\vfis{
V^{\rm{F}} \equiv e^{K}\Bigl(K^{,\alpha \bar{\beta}}D_{\alpha}W
\overline{D_{\beta}W}-3|W|^2\Bigr)} depends on the inflaton $\phi$
because $\phi$ necessarily appears in the K\"ahler potential. Even
if the superpotential depends weakly or not at all on $\phi$, the
total energy does vary with $\phi$.  Thus, restoring factors of
the Planck mass, we have \eqn\dvf{ V_{\rm{F}}'' =
{{K''\over{M_{\rm{p}}^2}}V_{\rm{F}} + ...}} and so a canonically-normalized
scalar has $\eta \sim 1$.  The only general solution to this
problem in F-term models is fine-tuning the contributions in \dvf\
to cancel each other to reasonable accuracy, leaving a small net
$\eta$.

\footline={\hss\tenrm\folio\hss}

    D-term models \BinetruyXJ, however, are well-known to be immune to the eta problem, as
the K\"ahler potential does not appear in the D-term energy.\foot{For a discussion of important updates to the D-term inflation scenario, see \BinetruyHH.}
This is argued to imply that the inflaton mass need not obey
$m_{\phi} \sim H$, as is generically true in F-term models, but
can instead be much smaller. This is a fairly strong argument in
favor of D-term inflation.

    The goal of this note is to demonstrate that this statement
no longer holds in string compactifications whose volume is
stabilized by a nonperturbative superpotential: both D-term and
F-term models, including shift-symmetric constructions,
receive inflaton mass corrections from threshold
corrections to the nonperturbative superpotential.  We will see
that these mass corrections are generically of order the Hubble
scale, so that $\eta \sim 1$.

   The source of the problem is readily understood.  Superpotential
stabilization of K\"ahler moduli proceeds by introducing an F-term
potential whose minimum determines\foot{In some cases, as we will
review, the physical volume and the stabilized K\"ahler modulus
are closely related but not identical.  For simplicity we will
nevertheless refer to this situation as `volume stabilization'.}
the compactification volume. Just as in the eta problem of F-term
inflation, this energy depends on the inflaton through the
K\"ahler potential. Although the inflationary dynamics may be
designed to proceed according to a weak interaction, e.g. of
widely-separated branes \refs{\dvalitye,\others}, the
inflaton-dependence of the volume-stabilizing F-term energy
typically introduces a {\it{stronger}} interaction and renders the
total potential too steep for inflation.\foot{This conflict
between F-term stabilization and slow-roll inflation was
recognized in a concrete form in the brane inflation
\refs{\dvalitye,\others} scenario of \kklmmt\ and has been
addressed in e.g.
\refs{\bhko,\bhk,\otherwarped,\shift,\pseudo,\BuchelQJ}.}

 A solution to this problem that has received considerable attention \refs{\shift,\watari,\angel,\sarah} is the introduction of continuous geometric symmetries to protect the inflaton mass.  In this approach, one posits
the existence of an approximate shift symmetry along the inflationary trajectory.

 One purpose of the present paper is to point out that one-loop threshold corrections to the volume-stabilizing nonperturbative superpotentials will typically lift any such shift symmetry and introduce an inflaton mass of order $H$.
Thus, shift symmetries do not suffice to protect the inflaton
mass, because quantum corrections will lift these symmetries and
change the inflaton potential.  Specifically, threshold
corrections to the nonperturbative superpotential introduce a
dependence of the F-term energy on the various moduli in the
system, including both open-string and closed-string fields.  The
inflaton is usually constructed as one of these moduli, so the
F-term potential depends on the inflaton. If this dependence is
generic then $\eta \sim 1$. This implies the existence of a rather
general eta problem for inflation in nonperturbatively-stabilized
string compactifications.

Volume stabilization is indispensable for a consistent model, and
at present the best-understood methods of volume stabilization use
nonperturbative superpotentials, along the lines suggested by
Kachru, Kallosh, Linde, and Trivedi (KKLT) \kklt.\foot{For a very
interesting example of perturbative volume stabilization, see
\alextwo.}  Thus, the presence of an eta problem in the context of
nonperturbative volume stabilization is an important aspect of
inflation in string theory.

We will be able to observe this effect in detail.  Berg, Haack,
and K\"ors (BHK) \refs{\bhko,\bhk} computed the one-loop threshold
corrections to the nonperturbative superpotential for the case of
type IIB string theory on certain toroidal orientifolds.  They
observed that the loop corrections introduce a moduli-dependent
mass for a mobile D3-brane in this background. (They further
showed that this mass correction may be used to fine-tune a
brane-antibrane potential to render it flat enough for inflation.)
Their result clearly demonstrates, for the case that the inflaton
is a D3-brane position and the compactification is a toroidal
orientifold, that the inflaton-dependence of the threshold
corrections is indeed sufficiently strong to affect inflation.

In \S5.2 we will apply the result of BHK to compute the inflaton
mass in the D3-D7 inflationary model \Kallosh.  A key point is
that the D3-D7 model is a D-term model that has been constructed
to enjoy a shift symmetry \shift, so it might be expected not to
be subject to an eta problem.  As we will see, even though D-term
inflation and shift symmetries do sometimes remove the usual eta
problem, neither one suffices to remove the eta problem explored
in this paper.

This statement should not be taken as a criticism of the D3-D7
model in particular.  We would expect similar results for nearly
any model of moving branes in a stabilized string
compactification.  More generally, the inflaton need not be a brane coordinate;
closed string moduli can certainly appear in the threshold
corrections, giving a mass to a closed string inflaton.  Moreover,
although nonperturbative superpotentials play an essential role in
our concrete discussion, any F-term moduli-stabilizing energy could in
principle lead to the same result.

\newsec{The Eta Problem in Supergravity}
In this section we will briefly review the supergravity eta
problem and mention how D-term models avoid the problem. In later
sections we will argue that this success of D-term models does
{\it{not}} extend to superpotential-stabilized string
compactifications.

\subsec{F-term Inflation and the Eta Problem }

In F-term models, inflation proceeds by slowly reducing the F-term energy,
\eqn\vfis{ V^{\rm{F}} \equiv e^{K}\Bigl(K^{,\alpha
\bar{\beta}}D_{\alpha}W \overline{D_{\beta}W}-3|W|^2\Bigr).}
    We are interested in computing the slow-roll parameter $\eta$
\etais.

Let us work with a canonically normalized inflaton $\phi$, which
we take to be complex for convenience. Then
$\partial_{\phi}\partial_{\bar{\phi}}K = 1$, so that as a function
of $\phi$, \eqn\phidep{ V^{\rm{F}}(\phi) = V^{\rm{F}}(0)
\Bigl(1+\phi\bar{\phi} + \ldots \Bigr)}

    We may therefore organize the contributions to \etaalso\ as
\eqn\dvftwo{ \eta = 1 + {e^{K}\over{V^{\rm{F}}(0)}}
\partial_{\phi}\partial_{\bar{\phi}}\Bigl(K^{,\alpha
\bar{\beta}}D_{\alpha}W \overline{D_{\beta}W}-3|W|^2\Bigr).}

    A successful model requires that the two terms on the right hand
side of \dvftwo\ are arranged to cancel to reasonable accuracy,
leaving a small net inflaton mass.  This sort of fine-tuning is
the only {\it{general}} solution to the $\eta$ problem in F-term
models.

    In particular, if the inflaton does not mix in the K\"ahler potential with any
other fields, so that $K_{,\alpha \bar{\phi}} = 0$ unless $\alpha
= \phi$, then the second term in \dvftwo\ depends on the inflaton
only through the superpotential, and the necessary fine-tuning
must be achieved by adjusting the inflaton-dependence of the
superpotential.

\subsec{D-term Inflation}

D-term models \refs{\BinetruyXJ,\Halyo} are those in which the
inflationary trajectory follows a direction which is not D-flat,
so that inflation proceeds by slowly reducing a D-term energy. The
particular advantage of this approach is that the K\"ahler
potential does not appear in the D-term energy, so the argument of
\S2.1 does not apply.  Thus, the inflaton mass does not receive
the corrections of order $H$ that plague F-term models.

At first sight, this conclusion appears surprisingly strong.  The mass terms given
in \S2.1 are merely a concrete example of a general expectation: because the inflationary energy
$V$ breaks supersymmetry, we expect soft scalar masses to be induced by gravitational
mediation, even if no more direct coupling is present.
The resulting masses will be of order $V/M_{\rm{p}}^2 = 3 H^2$.

Concretely, however, this problematic coupling of the inflaton to
the supersymmetry-breaking energy arises from the tree-level
K\"ahler potential for the case of F-term models.  D-term
inflation sidesteps the problem by providing an inflationary
energy which is insensitive to the K\"ahler potential
\dvali.\foot{S. Thomas has emphasized that Planck-suppressed
couplings of the inflaton in the K\"ahler potential can sometimes
produce an inflaton mass even in D-term models \refs{\dt,\scott}.}

We will find that this statement requires careful reexamination in
the context of stabilized string compactifications.  The reason is
that moduli stabilization typically introduces an F-term energy,
reviving the problem of \S2.1.

\newsec{Nonperturbative Superpotentials and Volume Stabilization}
The remainder of our discussion will rely on the details of moduli stabilization, so
in this section we will first outline the logic of moduli stabilization
and then explain how nonperturbative superpotentials can be used to fix
K\"ahler moduli.

\subsec{The Necessity of Volume Stabilization}

String compactifications on Calabi-Yau manifolds typically have a
large number of massless scalar fields, or moduli.  For our
purposes the most interesting moduli are the complex structure moduli, the positions of
D-branes, and the K\"ahler parameters, including the overall volume.

Moduli can ruin cosmological models in various ways.  They can
store energy during inflation and then interfere with
nucleosynthesis, or they could have time-dependent vevs at the
present epoch, leading to changes in various physical constants.
Finally, the presence of these light, gravitationally-coupled fields
would typically lead to unobserved fifth-force interactions.
Cosmological models which aim to be successful in detail should
somehow remove most or all of these light fields.

One modulus in particular presents a grave problem.  The overall
compactification volume does not have a flat potential, but is in
fact unstable: it has a runaway direction toward
decompactification. The reason is that the various sources of
inflationary energy in string theory will necessarily appear, in
the four-dimensional (Einstein-frame) description, multiplied by
inverse powers of the volume: \eqn\volprob{ V_{\rm{4d}} =
{C\over{\rho^{\alpha}}}.} Here $V_{\rm{4d}}$ is the inflationary
potential, $\rho$ is the volume modulus (taken to be real), $C$ is
a volume-independent factor, and $\alpha$ is positive. This result
is easily obtained by dimensional reduction of ten-dimensional
sources of energy, such as branes, strings, and fluxes.

If the volume were held fixed by hand, then a mild
inflaton-dependence in $C$ could lead to an inflating model.
However, in reality we expect that a fast roll in the $\rho$
direction, toward decompactification, will remove the possibility
of slow roll in the $\phi$ direction.

It is therefore absolutely essential to introduce some form of
volume-stabilizing potential $U(\rho)$, so that \eqn\vtot{ V =
{C\over{\rho^{\alpha}}} + U(\rho) } has a minimum at a finite
value of $\rho$.

The proposal of KKLT, which we will now review, is that a
nonperturbative superpotential could lead to the necessary
volume-dependence.

\subsec{Nonperturbative Superpotentials and Volume Stabilization}

Let us work in the concrete and well-studied example of the type
IIB string on a six-dimensional orientifold, which we view as a
limit of a compactification of F-theory on a fourfold. For
simplicity we assume that the threefold has exactly one K\"ahler
modulus, $\rho$.  Three-form fluxes $H_{3}, F_{3}$ in the internal
space lead to a superpotential \GVW\ \eqn\gvsup{ W_{0} = \int_{\rm{CY}}
(F_{3}-\tau H_{3})\wedge \Omega} which depends on the complex
structure moduli $\chi_{i}, i=0, \ldots h^{2,1}$ and the dilaton
$\tau$.

An additional contribution $W(\rho)$ to the superpotential would
allow simultaneous solution of \eqn\solve{ D_{\rho}W = D_{\tau}W =
D_{\chi_{i}}W =0.}  In this supersymmetric solution the dilaton,
the complex structure moduli, and the volume are stabilized.  (For
more details on the stabilization of the complex structure moduli
and the dilaton in this scenario, see e.g. \refs{\kklt, \GKP,
\PiSS}.)

KKLT proposed that a nonperturbative superpotential
$W_{{\rm{np}}}(\rho)$ from either of two sources could provide the
necessary effect:

(1) Euclidean D3-branes wrapping a divisor in the Calabi-Yau
\Witten.

(2) Gaugino condensation on a stack of $N>1$ D7-branes wrapping a
divisor in the Calabi-Yau, and filling spacetime.

In either case, the resulting superpotential takes the form

\eqn\bigwis{ W_{{\rm{np}}} = \Sigma(\zeta,\phi)e^{-a \rho}.} In
this formula $a$ is a numerical constant and $\Sigma$ is a
holomorphic function of the various moduli $\zeta$ (such as the
complex structure moduli $\chi_{i}$ and the positions of any
D-branes) and of the inflaton $\phi$.

In the absence of background flux, such a superpotential is possible only when
the divisor $D$ satisfies a rather stringent topological
condition: the arithmetic genus $\chi(D,{\cal{O}}_{D})$
of the divisor must obey $\chi=1$ \Witten.

As explained in \eucl, the effect of fluxes is to permit
gaugino condensation to occur somewhat more generally, so that divisors with
$\chi > 1$ can contribute to the superpotential.  There are reasons
to believe that the same conclusion applies to the Euclidean D3-brane superpotential \refs{\sandip,\sorokin}.

A special feature of the gaugino condensate superpotential is that
$a= 4\pi^2/N$ for the condensate of a pure $SU(N)$ gauge group,
whereas $a \sim 1$ for the case of Euclidean
D3-branes.\foot{Our conventions for $a$ and $\rho$ differ by a
factor of $(2\pi)$ from those of KKLT: $a_{KKLT} = 2\pi/N$.}

We will now turn our attention to the holomorphic prefactor
$\Sigma(\zeta,\phi)$.

\subsec{ Threshold Corrections to Nonperturbative Superpotentials}

Recall that in ${\cal{N}}=1$ Yang-Mills, the Wilsonian gauge
coupling is given by the real part of a holomorphic function $f$:
\eqn\act{ {1\over{g^2}} = {\rm{Re}}\Bigl(f(\zeta,\phi)\Bigr) }
This holomorphic coupling receives one-loop (and nonperturbative)
corrections, but no higher-loop corrections
\refs{\shifman,\NillesCY}, so that $f$ is the sum of a tree-level
piece and a one-loop correction: $f = f_0 + f_1$.

The one-loop correction $f_1$ is known as a ``threshold
correction" because it encodes the effect on the Wilsonian gauge
coupling of heavy particles at the threshold, i.e. at the
ultraviolet cutoff \louis.  This correction is a
holomorphic function of the moduli, including, in general, the inflaton.

The gaugino condensate superpotential in pure $SU(N)$ Yang-Mills with
ultraviolet cutoff $M_{\rm{UV}}$ and gauge kinetic function $f$ is
given by \shifman

\eqn\wgaug{ W = {16\pi^2}M_{\rm{UV}}^3
\exp\Bigl(-{8\pi^2\over{N}}f\Bigr) \equiv
\Sigma(\zeta,\phi)e^{-a\rho}.} We have absorbed the constants in
the exponent into $a$, we have omitted the dimensionful prefactor,
and we have used the fact that dimensional reduction of the $7+1$
dimensional theory on the D7-brane relates the tree-level gauge
coupling to the volume $\rho$ of the divisor.  All further moduli
dependence arising from $f_{1}$ has been encoded in
$\Sigma(\zeta,\phi)$.

In the remainder of the paper we will analyze the physical
consequences of the prefactor $\Sigma(\zeta,\phi)$, viewed as a
threshold correction to a gaugino condensate superpotential.  This
means that we are focusing our attention on gaugino condensation
instead of Euclidean D3-branes as the source of the
superpotential.

The motivation for this choice is that $\Sigma(\zeta,\phi)$ is
more readily computed in the gaugino condensate case.  For a
Euclidean D3-brane superpotential, $\Sigma(\zeta,\phi)$ represents
a one-loop determinant of fluctuations around the instanton.  In
the M-theory description of this effect, this depends on the
worldvolume theory of an M5-brane, which is rather subtle \Witten.
Although explicit results for $\Sigma$ are unavailable in the
Euclidean brane case, we do still expect to find nontrivial
inflaton-dependence, leading, as we will see for the gaugino
condensate case, to an eta problem.

\newsec{The Eta Problem in String Compactifications}

We will now examine the relation between moduli stabilization and
the eta problem.  In \S4.1 we recall a problem which can be
thought of as the incarnation of the (usual) supergravity eta
problem in a very specific string context.  Then, in \S4.2 we
explain how shift symmetries have been used to address this
problem, and we indicate a few important obstacles to the
construction of shift-symmetric models.

\subsec{Inflaton-Volume Mixing and the Eta Problem}

In the context of brane inflation in type IIB string theory, the
eta problem takes a novel form \kklmmt.  We will examine this now
because it presents a concrete setting in which shift symmetries
may be used to solve the usual eta problem.  Our eventual goal is
to understand a new and different eta problem which these
symmetries do {\it{not}} eliminate, but to achieve this it will be
very useful to review the shift symmetry idea in a simpler
setting.

D-brane inflation \dvalitye\ requires mobile, space-filling
D-branes, and in a type IIB compactification this is most simply
achieved with D3-branes.  It will be important for our
considerations that the coordinates of D3-branes (i.e., their
center-of-mass position moduli) $\phi_{i}, i=1,2,3$ appear in
the K\"ahler potential as \DeWolfe\foot{For additional explanation
of this point, see \kklmmt\ and especially \bhko.} \eqn\phikah{ K
= - 3 \log \Bigl(\rho + \bar{\rho} - k(\phi_i,\bar\phi_i)
\Bigr)} where $k(\phi_i,\bar\phi_i)$ is the (unknown)
K\"ahler potential for the Calabi-Yau manifold itself, which is
closely related to the D3-brane moduli space.  Singling out one
direction as the inflaton and denoting it by $\phi$, we have
$k(\phi,\bar\phi) = \phi\bar\phi + \ldots $, where the
expansion is performed around a point in the D3-brane moduli space
where the kinetic term is canonical.

This mixing of the brane coordinates with the geometric modulus
$\rho$ has important implications.  The physical volume $r$ in
this setting is no longer simply $ Re(\rho)$, but is instead
\eqn\volis{ 2 r = \rho + \bar{\rho} - \phi\bar\phi.}

This implies a revision of \volprob, namely
\eqn\newvolprob{V_{\rm{4d}} = {C\over{r^{\alpha}}} = {C\over{(\rho
-\phi\bar\phi/2)^{\alpha}}}} so that \kklmmt\ \eqn\visouch{
V_{\rm{4d}}(\phi) = V_{\rm{4d}}(0)
\Bigl(1+{\alpha\over{2r}}\phi\bar\phi \Bigr)}

This introduces a contribution of order one to $\eta$.  Because
this effect arises from a term in the K\"ahler potential, it is
reasonable to view it as the manifestation, in this specific model,
of the usual eta problem. (The new problem we will discuss shortly
does not have this property.)

\subsec{Solving the Eta Problem with Geometric Shift Symmetries}

Shift symmetries \shift\ are a promising approach to solving the
eta problem reviewed in the previous section.  The idea is to
consider a special compactification which happens to have a
particular continuous geometric symmetry.

The proposed symmetry is that the tree-level K\"ahler potential is
independent of one particular (real) field, such as the real part
of $\phi$.  There are strong arguments \refs{\angel,\watari}
from ${\cal{N}}=2$ gauged supergravity that this is indeed the
case in certain examples, at least before supersymmetry is broken.

The resulting K\"ahler potential, for example for D3-branes moving
along the torus directions of $K3 \times T^2$, takes the form
\eqn\shifty{ K = - 3 \log \Bigl( \rho + \bar\rho -
(\phi-\bar\phi)^2 \Bigr)} so that $Re(\phi)$ receives no
mass from the term analogous to \newvolprob.  This solves the eta
problem expressed in \visouch.

Various corrections\foot{Perturbative corrections
to the K\"ahler potential will almost certainly lift this
symmetry, although we will not address this
\refs{\BeckerNN,\forthcoming}.} will alter this result and lift
the shift symmetry of \shifty.  In particular, Berg, Haack, and
K\"ors have very clearly demonstrated that threshold corrections
to the D7-brane gauge coupling lift the shift symmetry of a
certain toroidal orientifold model.

We would like to observe that this conclusion is both generic and
problematic, and is in fact a symptom of a new eta problem for
string inflation.

Before moving to our main point, we pause to consider some of the
obstacles to implementing the shift symmetry argument. (This is an
aside because in \S5 we will ignore these difficulties and grant
the presence of such a symmetry, in the absence of threshold
corrections, and then demonstrate that the inclusion of threshold
corrections still causes an eta problem.)

The first difficulty is that requiring a geometric shift symmetry
places severe constraints on the compactification manifold.  It is
well-known (cf. \gsw, p.484) that ordinary Calabi-Yau threefolds, i.e.
Calabi-Yau threefolds whose holonomy is $SU(3)$ and not a
subgroup, do not have any continuous isometries. Thus,
orientifolds of tori and of $K3 \times T^{2}$ are the only
suitable candidates for shift-symmetric models. This implies a
tremendous reduction in the number of compactifications available
for model-building.

Furthermore, the strategy of guessing general results based on
detailed study of toroidal orientifold examples is not always
reliable.  In particular, even if most such simple examples have
continuous symmetries, we know for certain that ordinary
Calabi-Yau manifolds do not. Hence, any conclusions about shift
symmetries that are inferred from toroidal orientifold examples
apply only to that context, and not to the general case. This is
one of the reasons that our conclusions are different from those
of \sarah.

Moreover, some important aspects of model-building are actually
more difficult in the nominally simplified setting of toroidal
orientifolds.  Although partial stabilization of K\"ahler moduli
has been achieved in this context \refs{\kst,\tt}, complete
stabilization remains challenging.  At present it is not clear that known
methods will suffice to stabilize all the K\"ahler moduli in an
order-one fraction of toroidal orientifold models. In this regard,
Calabi-Yau threefolds with unreduced holonomy can be much more
tractable \DDF.  This is a fairly serious objection to toroidal
constructions, given the importance of moduli stabilization for an
inflationary model.  Even so, it is possible that complete moduli
stabilization will eventually be achieved for a toroidal
orientifold with properties appropriate for inflation.

\newsec{Threshold Corrections in Nonperturbative Superpotentials Change
the Inflaton Mass}

We now present the key observation of this paper, which is that
threshold corrections induce an entirely new eta problem which
D-term and shift-symmetry techniques do {\it{not}} solve.  That
is, we explain how threshold corrections lead to an inflaton mass
that is generically of order $H$, even in the special case that a
shift symmetry was present before the inclusion of these
corrections.

In \S5.1 we discuss the potential sources of an inflaton mass, and in
\S5.2 we illustrate our considerations with the D3-D7 model
\Kallosh, in which the problem is particularly clear.  In \S5.3 we
explore potential solutions to this problem.

\subsec{General Results}

The total potential in a stabilized inflationary model is the sum
of several contributions: \eqn\vtoa{ V
= V_{\rm{F}} + V_{\rm{pos}} + V_{\rm{int}} .}

The first contribution, $V_{\rm{F}}$, is the F-term moduli-stabilizing
energy. In the KKLT scenario, $V_{\rm{F}} = V_{\rm{AdS}} < 0$ is also the vacuum
energy of a supersymmetric $AdS_{4}$ solution.  A supersymmetry-breaking 
effect then adds an energy $V_{\rm{pos}}$ which `uplifts' the
total vacuum energy to a positive value, creating a metastable de
Sitter vacuum. The prototypical source of positive energy is an
anti-D3-brane \kklt, though there are various alternatives
\refs{\BKQ,\alexone}.

The final and most model-dependent ingredient is an interaction
potential $V_{\rm{int}}$ designed to produce the dynamics of slow-roll
inflation.  Simple examples include the weak interactions between
a widely-separated brane-antibrane pair \refs{\others,\kklmmt} or
between a D3-brane and a D7-brane \Kallosh.

The $\eta$ condition for slow-roll inflation (where primes denote
derivative with respect to the canonically-normalized inflaton) is
\eqn\fulleta{ V_{\rm{F}}'' + V_{\rm{pos}}'' + V_{\rm{int}}'' \ll 3H^2}

By far the simplest case has $V_{\rm{F}}$ and $V_{\rm{pos}}$ independent of
$\phi$, so that $\eta$ is determined by $V_{\rm{int}}''$ alone.  Then, if the
interaction potential is reasonably flat, the slow-roll condition
can be satisfied.  The only remaining challenge is to design an interaction
$V_{\rm{int}}(\phi)$ that is sufficiently weak.

Of course, this simple case is hard to achieve.  Let us now repeat
the potential problems:

(1) If $V_{\rm{int}}$ is an F-term energy then the $e^{K}$ prefactor
leads to an inflaton mass of order $H$.  This is the classic
supergravity eta problem \Copeland.

(2) If the inflaton and compactification volume mix, as in \volis,
and the energy is proportional to the volume, as in \newvolprob,
then this produces an eta problem as in \visouch.  This was the
problem in \kklmmt.

(3) If the volume-stabilizing $V_{\rm{F}}$ has inflaton dependence,
e.g. from threshold corrections, then this leads to yet another
eta problem.  The inflaton mass depends on the detailed form of
these threshold corrections, but is not expected to be
parametrically small.

D-term inflationary energy avoids the first problem, as we
recalled in \S2.2; shift symmetries \refs{\shift,\watari} avoid
the second problem, as we explained in \S4.2; but it appears that
some more clever mechanism, or an explicit fine-tuning, will be
necessary to overcome the third problem.  That is the point of the
present note.

\subsec{The Example of the D3-D7 Model}

It will be worthwhile to illustrate the assertions of the previous
section in a specific example. We will focus on the D3-D7 model of
\Kallosh.  This model is particularly interesting for our purposes
because it is a D-term model which can moreover be constructed to
take advantage of a shift symmetry, so that the first and second
problems of \S5.1 are not present. This leaves the inflaton mass
from threshold corrections as the final obstacle to a working
model.\foot{Berg, Haack, and K\"ors have done a careful study
\refs{\bhko,\bhk} of the inflaton mass corrections in the
brane-antibrane model of \kklmmt.  Because the second and third
effects listed in \S5.1 are {\it{both}} present in that example,
it is possible to balance these effects against each other and
fine-tune away the eta problem.  In contrast, our present point is
that the third effect, from threshold corrections, is problematic in
general, and particularly so in shift-symmetric models.}

We will now briefly review the aspects of the D3-D7 model
\Kallosh\foot{For a more recent generalization, see \refs{\DasguptaDW,\pisin}.}  that
are relevant for our considerations.  The general
proposal is that the weak interaction between a mobile D3-brane
and a D7-brane whose worldvolume flux ${\cal{F}}$ is not self-dual
can give rise to inflation.  The D3-brane moves toward the
D7-brane and then, at a critical distance, dissolves.

The flux in question is ${\cal{F}} \equiv dA - B$, where $A$ is
the gauge potential on the D7-brane worldvolume and $B$ is the
pullback of the NS-NS two-form potential.  If this flux is not
self-dual in the four-dimensional space described by the divisor
which the D7-brane wraps, then supersymmetry is broken and there
is a force between the D7-brane and the D3-brane \Kallosh.

This model can be compactified on $K3 \times T^{2}/{\Z_{2}}$, with
the orientifold action explained in \Kallosh.  Volume stabilization 
requires a stack of D7-branes wrapping the $K3$ and sitting at a particular
location on the torus, which we take to be the origin.  The D7-brane
bearing anti-self-dual flux may or may not sit at the same location.

Note that the translational symmetry along the torus may be
thought of as the origin of the shift symmetry \shift.
Correspondingly, the K\"ahler potential for this model is given by
the shift-symmetric form \shifty\refs{\shift,\angel,\watari}.

The holomorphic gauge coupling on the stack of D7-branes,
including the string loop correction, is \bhk\ \eqn\fis{ 2f = \rho
- {1\over{4\pi^2}} \log \vartheta_{1}(\phi,U) + \ldots }
where $\vartheta_{1}$ is a Jacobi theta function, $U$ is the complex
structure of the $T^{2}$, $\phi$ is the inflaton, and the omitted
terms are independent of $\phi$. Expanding, for convenience,
around $\phi = 1/2$, BHK find \eqn\wexp{
W_{{\rm{np}}}(1/2+\phi)=W_{{\rm{np}}}(1/2)\Big(1+
\delta(U)\phi^2\Bigr)} where \eqn\omegis{ \delta(U) =
{a\over{24}}\Bigl(E_{2}(U)+\vartheta_{3}(0,U)^4+\vartheta_{4}(0,U)^{4}\Bigr).}
Here $E_{2}$ is the second Eisenstein series, related to
derivatives of the $\vartheta$-functions, and $a$ is the numerical
constant appearing in \bigwis.  In these
expressions $\phi$ is dimensionless; the canonically-normalized
inflaton, with mass dimension one, is \eqn\varphiisnow{ \varphi =
M_{\rm{p}} \phi\sqrt{{3\over{\rho+\bar{\rho}}}}}.

We can now compute $\eta$ for the D3-D7 model on this
compactification, by using \wexp\ to expand the F-term energy. The
result, easily obtained using the SuperCosmology \supercosmo\
package, is conveniently expressed as \eqn\threemass{ \eta =
{4\over{3}}{\Bigl|{V_{{\rm{AdS}}}\over{V}}\Bigr|}\Bigl(2
{\delta(U)^2\over{a^2}}+ 3 {\delta(U)\over{a}}\Bigr)} where, as in
KKLT, $V_{{\rm{AdS}}}$ is the vacuum energy at the $AdS_{4}$
minimum which is uplifted to create a de Sitter vacuum.

This result is slightly different from the result of \bhk\ for the
mass of a $D3-\overline{D3}$ inflaton.  The reason is that the
K\"ahler potential relevant for brane-antibrane inflation is
\phikah, but for the present example of D3-D7 inflation the
K\"ahler potential takes the shift-symmetric form \shifty.

Let us now assess whether $\eta$ \threemass\ can satisfy the
slow-roll condition $\eta \le 10^{-2}$.  Each of the factors in
\threemass, except for $\delta(U)$, is roughly of order one or
larger.  The ratio $|V_{{\rm{AdS}}}|/V$ cannot be parametrically
small, because $V_{{\rm{AdS}}}$ determines the height of the
potential barrier that prevents decompactification, and the energy
density $V$ should not exceed this.\foot{This statement is
model-dependent; our present discussion assumes moduli
stabilization by the method of KKLT \kklt.} The constant $1/a$ is
likewise not parametrically small; in the concrete example given
in KKLT, $a$ was taken to be $2\pi/10$ (where we have included a
factor of $(2\pi)^{-1}$ which converts their result to our
notation), and more generally, $a= 4\pi^2/N$ for a stack of $N$
coincident D7-branes.

The only factor which might be small is $\delta(U)$.  As explained
in \bhk, $\delta$ is not automatically small, but there does exist
a small range of values of $U$, the torus complex structure, for
which $\delta(U) \ll 1$.  A small inflaton mass can therefore be
arranged by a choice of fluxes that fixes $U$ in this window. This
amounts to an explicit fine-tuning of the inflaton mass.

We conclude that the D3-D7 model requires a modest fine-tuning
which can be achieved by a judicious choice of fluxes.\foot{We
should again emphasize that corrections to the K\"ahler potential
may introduce further changes in the inflaton mass.}

\subsec{Discussion}

The result of the previous sections accords with the general
expectations discussed in \S2.  An inflaton mass which is much
smaller than $H$ does not arise automatically, nor even with the
imposition of a shift symmetry; in the end, a fine-tuning at the
percent level is necessary to make the model work.  In the scheme
of inflationary fine-tuning, this is not a serious problem; in
particular, it should be contrasted to the functional fine-tuning
required for certain models in which $\phi \gg M_{\rm{p}}$.  Even so,
the necessity of fine-tuning in the present case cannot be
ignored.

This result should not be interpreted as a stroke against the
D3-D7 model (or any other model) in particular.  In fact, we would
expect almost any complete and fully-realized model to require
some fine-tuning of parameters.  Omission or simplification of
certain physical ingredients, especially moduli stabilization, may
obscure the eta problem and make a model appear to work
automatically, but sufficient inspection can be expected to reveal
one or more problems of detail that require fine-tuning.

It would be extremely interesting to find a solution to this eta
problem that does not amount to a fine-tuning of parameters.  A
slightly modified mechanism of volume stabilization, such as the
proposal of \kl, does alter the mass formula \threemass, but does
not naturally produce a small mass.  However, it may be possible
to invent a method of volume stabilization which does not affect
the inflaton mass.  Volume dependence through a D-term energy
would be a promising candidate.

Another interesting possibility \dvali\ is that an inflaton
charged under a symmetry $G$ can sometimes be excluded from the
holomorphic correction term $f_{1}$, so that
${\partial{f_{1}}\over{\partial{\phi}}}=0$. However, in simple
examples, such as the D3-D7 model, no such symmetry is present.
Moreover, D-term inflation requires \BinetruyXJ\ that $\phi$ is
neutral under the $U(1)$ gauge group $G_{D}$ whose D-term energy
drives inflation, so in particular $G$ cannot coincide with
$G_{D}$.  It is reasonable to expect, however, that discrete
symmetries of the appropriate form can sometimes be arranged.

\newsec{Conclusion}

We have seen that threshold corrections to volume-stabilizing
nonperturbative superpotentials create an eta problem for
inflationary models in string theory.  These threshold corrections
cause the volume-stabilizing F-term energy to depend, generically,
on the values of the open-string and compactification moduli.
Because the inflaton is expected to consist of one of these
moduli, the threshold correction changes the dependence of the
inflationary energy on the inflaton vev, altering the slow-roll
parameters and creating an eta problem.

This conclusion applies to models which satisfy several
assumptions, which we now repeat for clarity.  Our general
considerations were limited to models of inflation which can be
realized in a string compactification. In any such model it is
essential that the instability to decompactification has been
removed by moduli stabilization; it is also desirable that all
other moduli have also been stabilized.  We have explicitly
assumed that the volume stabilization arises from a
nonperturbative contribution to the superpotential, as in KKLT
\kklt. (For interesting alternatives, see \alextwo.) We have
also assumed that the inflaton is a modulus whose flat
direction is slightly lifted by a further supersymmetry-breaking effect.
This could correspond, for example, to a brane interaction.

Thus, our result applies to any model of inflation in string
theory which uses a compactification stabilized by methods
analogous to those of KKLT.  Every aspect of the discussion is
simplest in the case of D-brane inflation in a type IIB
compactification, but the result applies much more broadly.  For
example, current techniques for moduli stabilization in the
heterotic string \refs{\hetero} and in M-theory on $G_{2}$
manifolds \lukas\ also use a combination of flux and
nonperturbative superpotentials.  Any inflationary model\foot{For
interesting examples in this category, see \hetinf.} which is
elaborated on one of these foundations would be subject to an eta
problem from threshold corrections to these superpotentials.

Moreover, although we have seen that the threshold corrections of
Berg, Haack and K\"ors \bhko\ lead to an explicit result for the
inflaton mass in the D3-D7 model on $K3 \times T^{2}$, generic
moduli dependence will lead to an eta problem even in more
complicated cases.  For example, the threshold corrections are not
known for generic Calabi-Yau threefolds, so no complete and
explicit computation of the slow-roll parameters is possible at
present for an inflationary model arising in a compactification on
such a space.  Progress in this direction appears to be important
for inflationary model-building in string compactifications.

It is essential to recognize that although the conclusions of this
paper are somewhat general, the actual computation of the inflaton mass
is only strictly applicable to a supersymmetric $AdS_{4}$ configuration
that can be uplifted to produce an inflationary scenario.  In
particular, the one-loop exactness of threshold corrections in
supersymmetric theories permits us to be somewhat precise about the inflaton
mass in a supersymmetric vacuum, but, as we have emphasized throughout,
supersymmetry-breaking effects will typically produce substantial corrections to these mass terms.

Nevertheless, the strategy of understanding the lifting of (inflaton) flat
directions in a supersymmetric vacuum is a sensible one.\foot{This
perspective was the one used to expose and address the problem of a brane-antibrane inflaton mass in \refs{\kklmmt,\otherwarped,\shift}.}  If no
suitably flat direction exists in the supersymmetric configuration, it is
very hard to believe that the addition of gravitationally-mediated soft terms will
remedy this problem.  Moreover, it is usually not possible to compute
these corrected masses in detail.

Thus, it is usually impossible to prove that a given string model
has a small inflaton mass, including all quantum corrections.  On
the other hand, it is possible to establish that a given
model has an eta problem, because if a problem arises from one set
of quantum corrections, such as threshold corrections to the gauge
coupling, then further quantum corrections will generically not
undo this problem. In this paper we have focused on establishing a
problem using the one-loop-exact results for the superpotential,
with the understanding that additional corrections, e.g. to the
K\"ahler potential, should not conspire to flatten the inflaton
potential.

There are several interesting directions for future work. First of
all, it is the threshold corrections from closed string moduli
that are relevant when the inflaton itself is a closed-string
field, for example a geometric modulus \eight.  The mass of such a
closed string inflaton depends on these corrections, and it would
be useful to understand their form.

Furthermore, we have only examined the nonperturbative
superpotentials resulting from gaugino condensation, but Euclidean
D3-branes are known to play an important role in stabilizing
certain classes of K\"ahler moduli \refs{\eucl,\DDF}.  In this
context the inflaton dependence of the instanton superpotential
arises through a moduli-dependent one-loop determinant
$\Sigma(\zeta,\phi)$ of fluctuations around the instanton.  It
would be extremely interesting, although challenging \Witten, to
compute prefactors of this sort, not only for the considerations
of this paper, but for rather general moduli stabilization.

In addition, corrections to the K\"ahler potential can further
adjust the dependence of the total inflationary energy on the
inflaton vev. A complete and consistent model requires inclusion
of these effects, which have also not yet been calcuated.

Looking forward, we can hope that a thorough understanding of
the effect of threshold corrections on shift-symmetric brane configurations will
guide us to models in which the threshold corrections, and all
other quantum corrections, are indeed small, so that the shift
symmetry is an approximate symmetry of the full quantum theory. If
this could be achieved, it would be a significant step toward a
controllable model of inflation.

\beginsection{ Acknowledgements}

I would like to thank Cliff Burgess, Gia Dvali, Jonathan Hsu, Renata Kallosh,
Eva Silverstein, Scott Thomas, Sandip Trivedi, and Marco Zagermann
for useful comments. I am especially indebted to Marcus Berg,
Michael Haack, and Shamit Kachru for extensive discussions and for
comments on the manuscript. Furthermore, I am grateful to the
Perimeter Institute and the Tata Institute for Fundamental
Research for hospitality during the completion of this work, and
to the organizers of the Fifth PI/UT/CITA Workshop and of the
Indian Strings Meeting 2004 for the opportunity to present this
material. This research was supported in part by the Department of
Energy under contract number DE-AC03-76SF00515.

\vfil\eject\appendix{A}{A Field-Theory Model of the Brane Interaction}

In this appendix we will point out a counterintuitive aspect of our conclusion.  We will then 
use a field-theory model to expose the flaw in this intuition, and to 
further demonstrate that our results are correct.

The inflaton mass term from threshold corrections is the result of an interaction induced by massive strings stretched between the D3-brane and the
D7-branes, which we refer to as 3-7 strings.  In the field theory description, these 3-7 strings correspond to a massive flavor whose mass $m_{37}$ is
controlled by the modulus $\phi$.  From this perspective, one might expect this massive flavor to decouple when its mass is very large, and to give
rise to a negligible interaction in that limit.  It is therefore somewhat surprising that the inflaton mass \threemass\ does not diminish when $|\phi|$
is large.  Should we not expect the BHK result to vanish for widely-separated branes?

To resolve this puzzle, we first note that as soon as $m_{37}$ approaches the mass of a string winding the torus,
the D3-D7 interaction induced by the superpotential is correctly described by the full string threshold correction of BHK, and not by its field theory limit.  Thus, we can place an upper limit $ \Lambda_{\rm{UV}} < m_{\rm{W}} $ on the ultraviolet cutoff of our field-theory description, where $m_{\rm{W}}$ denotes the mass of the lightest wound string.
In other words, the field theory that provided the decoupling intuition applies only to situations in which the brane separation is much less than the smallest radius of the torus.\foot{I am grateful to M. Berg and M. Haack for discussions on this point.}  At greater separations, wound strings can appear in the theory and contribute an additional interaction between the D3-brane and the D7-branes.

We should therefore ask whether decoupling sufficient for slow roll is possible within this limit imposed by the radius of the compact space.  To do this, we will examine a simple field theory that models the D3-D7 interaction induced by stretched (but not wound) 3-7 strings. (We will check our model by verifying that it coincides with the small-separation limit of the full BHK result.)

The model is a supersymmetric $SU(N)$ Yang-Mills theory 
with a single chiral superfield $Q$ whose mass
is controlled by a parameter $\phi$.  Here we will take $\phi$ to
be non-dynamical, and will examine the gaugino condensate
superpotential as a function of $\phi$.\foot{For simplicity we are
studying the supersymmetric configuration; the
supersymmetry-breaking effects used in the model of \Kallosh\
would generate additional corrections to the inflaton mass, in
addition to introducing a tachyon.}

The gaugino condensate superpotential below the scale $m_{37}$, i.e. after integrating out $Q$,
can be matched to the superpotential above this scale.
For $N >2$ the result is simply \intriligator\ \eqn\wlow{ W_{{\rm{low}}} \propto
\Lambda_{{\rm{high}}}^{3-1/N} m_{37}^{1/N} } with $\Lambda_{\rm{high}}$ 
the dynamically-generated scale of the high-energy theory.
Thus, in the low-energy theory,
$W = C \phi^{1/N}$ with $C$ independent of $\phi$.  We can
precisely reproduce \wlow\ by expanding the
superpotential \wgaug, including the full string threshold correction \fis\
of BHK, in the limit $\phi \ll 1$, after using the relation $a = 4\pi^2/N$.

Let us now compute $\eta$ in this model.  From the supergravity formula for the F-term energy, we have\eqn\vlow{ V = - 3 e^{K} C^2 \phi^{2/N} } so
that\foot{We have again replaced the dimensionless $\phi$ with the
canonically-normalized $\varphi$, cf. \varphiisnow.} \eqn\etalow{
\eta = -{2\over{N}}\left(1-{2\over{N}}\right)\left({M_{\rm{p}}\over{\varphi}}\right)^2.}
Taking $\rho$ real and using $\varphi = M_{\rm{p}}\phi
\sqrt{{3\over{2\rho}}}$, we have \eqn\etasimpl{ \eta \sim
-{2\over{N}}\left({{2\rho}\over{3\phi^2}}\right)} so that applying
$a = 4\pi^2/N$, we finally come to \eqn\etafinal{ \eta \sim -{{ a
\rho }\over{3\pi^2\phi^2}}.} However, $a\rho \gg 1$ was a
condition for the validity of the nonperturbative superpotential
used by KKLT: \bigwis\ is the leading approximation, analogous to
a single-instanton effect, and there will be corrections
suppressed by further powers of $e^{-a\rho}$.  Furthermore, $|\phi| \le {1\over{2}}$ measures the distance from the origin on a unit torus, so $|\phi| \ll {1\over{2}}$ is necessary in order for the brane separation to be small compared to the size of the torus (and hence for the field theory model to be a good approximation to the true result, which incorporates wound strings.)  Thus, there
is no controllable parameter regime in which \etafinal\ is small.

Indeed, even at the extreme boundary of the region of control, $a\rho \sim 1, |\phi| \sim {1\over{2}}$, we have at best
$\eta \sim {1\over{7}}$.  If we were to extend the toy model to $\varphi > M_{\rm{p}}$ then the
interaction would no longer be strong enough to affect slow-roll.  However, this is not an allowed range in the full
model, because of the UV cutoff of the effective field
theory, which corresponds to the limit imposed by the radius of the
compact space.

We conclude that one cannot arrange suitable decoupling simply by separating the branes;
 a somewhat more complicated fine-tuning will be necessary to
remove the inflaton mass terms under consideration.  We have certainly not demonstrated
that slow-roll is impossible for D3-D7 systems in the regime in
which separations are small compared to the size of the torus.  We have simply shown that the
interaction captured by threshold corrections produces, on its
own, an unsuitably large inflaton mass in this range, so that some fine-tuning against
other effects would be needed to make a phenomenologically
acceptable model.  Thus, one cannot evade the arguments of this paper by separating the
D3-brane from the D7-branes and invoking decoupling.

\listrefs
\end